# Thermopower modulation clarification of the intrinsic effective mass in a transparent oxide semiconductor, BaSnO$_3$

Anup V. Sanchela[1,*], Takaki Onozato[2], Bin Feng[3], Yuichi Ikuhara[3] and Hiromichi Ohta[1,*]

[1]*Research Institute for Electronic Science, Hokkaido University, N20W10, Kita, Sapporo 001-0020, Japan*

[2]*Graduate School of Information Science and Technology, Hokkaido University, N14W9, Kita, Sapporo 060-0814, Japan*

[3]*Institute of Engineering Innovation, The University of Tokyo, 2-11-16 Yayoi, Bunkyo, Tokyo 113-8656, Japan*

[*]To whom all correspondence should be addressed: anup.sanchela@es.hokudai.ac.jp, hiromichi.ohta@es.hokudai.ac.jp

Subject Areas: Condensed Matter Physics, Materials Science

**Although there are so many reports on the carrier effective mass ($m^*$) of a transparent oxide semiconductor BaSnO$_3$, it is almost impossible to know the intrinsic $m^*$ value because the reported $m^*$ values are scattered from 0.06 to 3.7 $m_0$. Here we successfully clarified the intrinsic $m^*$ of BaSnO$_3$, $m^*$=0.40±0.01 $m_0$, by the thermopower modulation clarification method. We also found the threshold of degenerate/non-degenerate semiconductor of BaSnO$_3$; At the threshold, the thermopower value of both La-doped BaSnO$_3$ and BaSnO$_3$ TFT structure was 240 μV K$^{-1}$, bulk carrier concentration was 1.4×10$^{19}$ cm$^{-3}$, and two-dimensional sheet carrier concentration was 1.8×10$^{12}$ cm$^{-2}$. When the $E_F$ locates above the parabolic shaped conduction band bottom, rather high mobility was observed. On the contrary, very low carrier mobility was observed when the $E_F$ lays below the threshold, most likely due to that the tail states suppress the carrier mobility. The present results are useful for further development of BaSnO$_3$ based oxide electronics.**

Today, transparent oxide semiconductors (TOSs), which show rather high electrical conductivity with large bandgap ($E_g > 3.1$ eV), including Sn-doped $In_2O_3$ (ITO) and $InGaZnO_4$-based oxides are widely applied as transparent electrodes and channel semiconductors for thin-film transistor (TFT) driven flat panel displays (FPDs) such as liquid crystal displays (LCDs) and organic light emitting diodes (OLEDs) [1]. Since the performance of TFT strongly depends on the carrier mobility of channel semiconductor, novel TOSs exhibiting higher carrier mobility have been intensively explored thus far. In 2012, Kim *et al*. reported that La-doped $BaSnO_3$ (Space group: $Pm\bar{3}m$, Cubic perovskite structure, $a$=4.115 Å, $E_g \sim 3.1$ eV) single crystal, which was grown by the flux method, exhibited very high mobility ($\mu_{Hall} \sim 320$ cm$^2$ V$^{-1}$ s$^{-1}$) at room temperature [2, 3]. This report ignited a fire of current research on $BaSnO_3$ films and $BaSnO_3$ based TFTs [4-9].

Since the mobility is expressed as $\mu = e \cdot \tau \cdot m^{*-1}$, where $e$, $\tau$, and $m^*$ are electron charge, carrier relaxation time, and carrier effective mass, the high mobility of the La-doped $BaSnO_3$ single crystal would be due to both small $m^*$ and large $\tau$. Generally, $\tau$ value of epitaxial films is smaller than that of the bulk single crystal due to the fact that the carrier electrons are scattered at dislocations, which are originated from the lattice mismatch ($\delta$). Since the $\delta$ between $BaSnO_3$ and $SrTiO_3$ ($a$=3.905 Å) is +5.3 %, a misfit dislocation spacing ($d$) of 7.4 nm can be estimated; $BaSnO_3$ films grown on (001) $SrTiO_3$ substrates exhibited rather small mobilities ($\mu_{Hall}$ 26−100 cm$^2$ V$^{-1}$ s$^{-1}$) [3, 7, 10] as compared with that grown on (001) $PrScO_3$ ($a$=4.026 Å, $\delta$=+2.2 %, $d$ ~17.7 nm) ($\mu_{Hall}$ 150 cm$^2$ V$^{-1}$ s$^{-1}$) [7]. On the other hand, $m^*$ only depends on the electronic structure of the material. Although there are many theoretical and experimental values (mostly determined from the optical properties) of $m^*$ have been reported, e.g. ~0.06 $m_0$ [11], ~0.4 $m_0$ [3], and ~0.2 $m_0$ [12] for theoretical and 3.7 $m_0$ [13], 0.61 $m_0$ [14], ~0.35 $m_0$ [15], ~0.396 $m_0$ [16], 0.27±0.05 $m_0$ [10], 0.19±0.01 $m_0$ [8] for experimental, it is almost impossible to choose the intrinsic $m^*$ value.

We speculated that the experimental values include some errors since the substrate contributes the optical spectra of the $BaSnO_3$ films. In order to overcome this difficulty, we used thermopower ($S$) modulation clarification method to determine the intrinsic $m^*$ of BaSnO$_3$ since $S$ clearly reflects the energy derivative of the density of states (DOS) at the Fermi energy ($E_F$). In this study, we measured $S$ of La-doped BaSnO$_3$ films with varied carrier concentration and BaSnO$_3$ with TFT structure, in which carrier concentration can be modulated by applying the electric field. As a result of this study, we successfully clarified the intrinsic $m^*=0.4\pm0.01\ m_0$ when the $E_F$ locates above the degenerate/non-degenerate threshold of BaSnO$_3$.

La-doped BaSnO$_3$ films (Thickness: ~200 nm) were heteroepitaxially grown on (001) SrTiO$_3$ single crystal substrates by pulsed laser deposition (PLD, KrF excimer laser, $\lambda$=248 nm, 10 Hz, Fluence ~2 J cm$^{-2}$ pulse$^{-1}$) technique using dense ceramic disks of Ba$_{1-x}$La$_x$SnO$_3$ (0.001≤ $x$ ≤ 0.07) as the targets. During the film growth, the substrate temperature and oxygen pressure were kept at 700 °C and 10 Pa, respectively. Since the surface of as-deposited BaSnO$_3$ films were composed of tiny (~50 nm) grains, we annealed the film at 1200 °C in air to obtain atomically smooth surface [17]. High-resolution X-ray diffraction (Cu K$\alpha_1$, ATX-G, Rigaku Co.) measurements revealed that the resultant films were heteroepitaxially grown on (001) SrTiO$_3$ substrate with cube-on-cube epitaxial relationship. The film thickness of the resultant films was calculated from the Pendellösung fringes, which was observed around 002 diffraction peak of La-doped BaSnO$_3$. Stepped and terraced surface was observed in the topographic AFM image of the resultant films.

We then measured the electrical conductivity ($\sigma$), carrier concentration ($n$), and Hall mobility ($\mu_{Hall}$) of the La-doped BaSnO$_3$ films at room temperature by a conventional d.c. four-probe method, using an In–Ga alloy electrode with van der Pauw geometry. $S$ was measured at room temperature by creating a temperature gradient ($\Delta T$) of ~4 K across the film using two Peltier devices (while using two small thermocouples to monitor the actual temperatures of each end of the Ba$_{1-x}$La$_x$SnO$_3$ films). The thermo-electromotive force ($\Delta V$) and $\Delta T$ were measured simultaneously, and $S$-values were obtained from the slope of the $\Delta V$–$\Delta T$ plots. **Table I** summarizes the electrical properties of the La-doped BaSnO$_3$ films at room temperature. The $\sigma$ rapidly increased with $x$ in $Ba_{1-x}La_xSnO_3$ and reached maximum value of 6910 S cm$^{-1}$ for $x=0.05$ sample. At the same time, $n$ and $\mu_{Hall}$ achieved highest value of $6.41\times10^{20}$ cm$^{-3}$ and 67.3 cm$^2$ V$^{-1}$ s$^{-1}$. It should be noticed that the $\mu_{Hall}$ value increased with increasing $n$, consistent with the data reported by Kim et al.[2] The sign of $S$ was negative in all samples, consistent with the fact that La-doped $BaSnO_3$ is an n-type semiconductor. The absolute value $|S|$ gradually decreased with $n$.

In order to clarify the intrinsic $m^*$, we plotted the relationship between $n$ and $S$. We were then able to calculate the DOS effective mass ($m^*$), as outlined in equations (1)–(3) below. When $n$ exceeds ~$10^{19}$ cm$^{-3}$, $|S|$ was found to monotonically decrease with increasing $n$, as shown in **Fig. 1**. We then calculated $m^*$ using the following relation between $n$ and $S$ [18],

$$S = -\frac{k_B}{e}\left(\frac{(r+2)F_{r+1}(\xi)}{(r+1)F_r(\xi)} - \xi\right) \quad (1)$$

where $k_B$, $\xi$, $r$, and $F_r$ are the Boltzmann constant, chemical potential, scattering parameter of relaxation time, and Fermi integral, respectively. $F_r$ is given by,

$$F_r(\xi) = \int_0^\infty \frac{x^r}{1+e^{x-\xi}}dx \quad (2)$$

and $n$ by,

$$n_- = 4\pi\left(\frac{2m^*k_BT}{h^2}\right)^{3/2} F_{1/2}(\xi) \quad (3)$$

where $h$ and $T$ are the Planck constant and absolute temperature, respectively. Using the equations (1)–(3), we successfully obtained the intrinsic $m^*=0.40\pm0.01$ $m_0$ when $n \geq 1.4\times10^{19}$ cm$^{-3}$, though the $S-n$ plots at $n < 1.4\times10^{19}$ cm$^{-3}$ did not obey the $m^*=0.4$ $m_0$ line. We judged there is a threshold of degenerate/non-degenerate semiconductor at around $n=1.4\times10^{19}$ cm$^{-3}$. As shown in the inset, when the $n$ exceeds the threshold (≡Degenerate semiconductor), the $E_F$ locates inside of the parabolic shaped conduction band. On the other hand, when the $n$ is less than the threshold (≡Non-degenerate semiconductor), the $E_F$ locates at the non-parabolic shaped extra DOS. At the threshold, $S$ value should be 240 µV K$^{-1}$.

In order to further clarify the intrinsic $m^*$ of $BaSnO_3$, we measured electric field modulated $S$ of $BaSnO_3$ based TFT as shown in **Fig. 2**. First, 22-nm-thick undoped $BaSnO_3$ film was prepared by PLD as described above. Stepped (~0.4 nm) and terraced surface was clearly observed as shown in **Fig. 2(d)**. Then, 30-nm-thick Ti films were deposited with stencil mask by electron beam (EB; base pressure ~$10^{-4}$, no substrate heating) evaporation, which were used as a source and drain electrode. The channel length ($L$) and the channel width ($W$) of the TFT were 800 μm and 400 μm respectively. After that, ~300-nm-thick amorphous 12CaO·7Al$_2$O$_3$ ($a$-C12A7, $\varepsilon_r$=12) [19] film used as a gate insulator which was deposited through a stencil mask by a PLD. Finally, for top gate electrode, ~30-nm-thick Ti film was deposited by EB evaporation as mentioned above. As fabricated TFT device was annealed at 150 °C in air to reduce the off current. Detail of our TFT fabrication process is described elsewhere [19-23]. **Figure 2(e)** shows the cross sectional high angle annular dark field scanning transmission electron microscope (HAADF-STEM, JEM-ARM200F with an accelerating voltage of 200 kV) image of the Ti/$a$-C12A7/BaSnO$_3$/SrTiO$_3$ interfacial region. We clearly observe a multilayer structure. The thicknesses of Ti, $a$-C12A7 and BaSnO$_3$ are 28, 280 and 22 nm, respectively. Abrupt heterointerface between $a$-C12A7/BaSnO$_3$ is clearly seen.

The semiconductor device analyzer (Agilent B1500A) was used to measure the transistor characteristics of undoped BaSnO$_3$ TFT at room temperature. **Figure 3** summarizes typical transistor characteristics such as transfer and output characteristics, threshold voltage, field effect mobility of the resultant BaSnO$_3$ TFT. The drain current ($I_d$) increases markedly as gate voltage ($V_g$) increased **[Fig. 3(a)]**, which indicates the channel was n-type and electron carriers were accumulated by positive $V_g$. An on-to-off current ratio ~$10^3$ was obtained for $V_d$= 1 V. We observed rather large threshold voltage ($V_{th}$) of +5.5 V, which was obtained by the linear fit of $I_d^{0.5}$−$V_g$ plot **[Fig. 3(b)]**. Clear pinch-off behavior and current saturation in the $I_d$ reveal that the resultant TFT obeys with the standard field effect transistor theory **[Fig. 3(c)]**. The field effect mobility ($\mu_{FE}$) was calculated from $\mu_{FE}=g_m[(W/L)C_i \cdot V_d]^{-1}$, where $g_m$ is transconductance $\partial I_d/\partial V_g$ and $C_i$ is the capacitance per unit area (39 nF cm$^{-2}$) **[Fig. 3(d)]**. The $\mu_{FE}$ drastically increases with $V_g$ and saturates at ~40 cm$^2$ V$^{-1}$ s$^{-1}$, which is ~60 % of the room temperature $\mu_{Hall}$ of La-doped BaSnO$_3$ ($\mu_{Hall}$ ~67 cm$^2$ V$^{-1}$ s$^{-1}$).

Then we measured electric field modulated $S$ of the resultant BaSnO$_3$ TFT **(Fig. 2)**. For the $S$ measurements, we used two Peltier devices, which were placed under the TFT, to give a temperature difference between the source and drain electrodes **[Fig. 2(b)]**. Two thermocouples (K-type, 150 μm in diameter, SHINNETSU Co.), which were mechanically attached at both edges of the channel, monitored the temperature difference ($\Delta T$, 0–5 K) **[Fig. 2(c)]**. The thermo-electromotive force ($\Delta V$) and $\Delta T$ values were simultaneously measured at room temperature. The $S$-values were obtained from the slope of the $\Delta V$–$\Delta T$ plots. Detail of our electric field modulated $S$ measurement is described elsewhere [19, 20, 22-25].

**Figure 4** shows electric field modulated $S$ of the resultant BaSnO$_3$ TFT as a function of sheet carrier concentration ($n_s$), which was deduced from $n_s = C_i \cdot (V_g - V_{th})$. The $|S|$ value gradually decreases from 308 to 120 μVK$^{-1}$ with $n_s$, consistent with the La-doped BaSnO$_3$ films **(Fig. 1)**. It should be noted that there is a deflection point at around ($|S|$, $n_s$)=(240 μV K$^{-1}$, 1.8×10$^{12}$ cm$^{-2}$). Almost linear relationship can be observed in the $S$−log $n_s$ plot when $n_s$ exceeds 1.8×10$^{12}$ cm$^{-2}$, the slope is ~200 μV K$^{-1}$/decade while that is not straight line when $n_s$ below 1.8×10$^{12}$ cm$^{-2}$. This observation is basically similar to that of the La-doped BaSnO$_3$ films; Degenerate/non-degenerate threshold is located at around ($|S|$, $n_s$)=(240 μV K$^{-1}$, 1.8×10$^{12}$ cm$^{-2}$). In the degenerate region, the $E_F$ locates above the threshold and $S$ obeys the Eq. (1)−(3); The conduction band bottom is parabolic shaped and the intrinsic $m^*$ is 0.40±0.01 $m_0$.

In the non-degenerate region, $E_F$ lays below the threshold. $S$ value does not show clear $n$ dependence. Similar phenomena were also observed in amorphous TOSs, $a$-InGaZnO$_4$ and $a$-In$_2$MgO$_4$ [22]. In case of these amorphous TOSs, structural imperfections lead to the formation of antiparabolic shaped extra state just below the original conduction band bottom and plays an essential role in transistor switching of TOS-based TFTs [26, 27]. In the present case, almost linear shaped extra DOS i.e. tail states would be formed just below the conduction band bottom, possibly due to oxygen deficiency. Such tail states suppress the carrier mobility, therefore,

BaSnO$_3$ films exhibited very low mobility when the $E_F$ laid below the threshold.

In summary, we have clarified the intrinsic effective mass of a transparent oxide semiconductor BaSnO$_3$, $m^*$=0.40±0.01 $m_0$, by the thermopower modulation clarification method. We also found the threshold of degenerate/non-degenerate semiconductor of BaSnO$_3$; At the threshold, the thermopower value of both La-doped BaSnO$_3$ and BaSnO$_3$ TFT structure was 240 μV K$^{-1}$, bulk carrier concentration was $1.4 \times 10^{19}$ cm$^{-3}$, and two-dimensional sheet carrier concentration was $1.8 \times 10^{12}$ cm$^{-2}$. When the $E_F$ locates above the parabolic shaped conduction band bottom, rather high mobility was observed. On the contrary, very low carrier mobility was observed when the $E_F$ lays below the threshold, most likely due to that the tail states suppress the carrier mobility.

We hope the present results are useful for further development of BaSnO$_3$ based oxide electronics.

**ACKNOWLEDGEMENTS**

This research was supported by Grants-in-Aid for Scientific Research on Innovative Areas "Nano Informatics" (25106003 and 25106007) from the Japan Society for the Promotion of Science (JSPS). H.O. was supported by Grants-in-Aid for Scientific Research A (17H01314) and B (26287064) from the JSPS and the Asahi Glass Foundation. A part of this work was supported by "Nanotechnology Platform" (12024046), sponsored by MEXT, Japan.

**FIGURE CAPTIONS**

**FIG. 1**| Change in the thermopower ($S$) of the La-doped BaSnO$_3$ films grown on (001) SrTiO$_3$ substrate as a function of the carrier concentration ($n$) at room temperature. The red line is a guide to the eye. The effective mass ($m^*$) values, calculated using the Eq. (1)−(3) are shown. The gray line is the $S$−$n$ curve, which was calculated using $m^*$=0.4±0.1 $m_0$. The dotted line is a threshold of degenerate/non-degenerate semiconductor at around $n$=1.4×10$^{19}$ cm$^{-3}$. The inset

shows the schematic explanation of energy dependence of the DOS.

**FIG. 2|** The BaSnO$_3$ TFT structure [(a) Schematic structure, (b) photograph of the thermopower measurement, (c) the magnified photograph, (d) topographic AFM image of the BaSnO$_3$ film surface, and (e) cross-sectional HAADF-STEM images]. (a-c) The channel length ($L$) and the channel width ($W$) of the TFT were 800 μm and 400 μm, respectively. The capacitance per unit area ($C_i$) of the $a$-C12A7 gate insulator is 39 nF cm$^{-2}$. Two thermocouples (K-type, 150 μm in diameter, SHINNETSU Co.), which were mechanically attached at both edges of the channel, monitored the temperature difference. Stepped and terraced surface is seen in (d). (e) Multilayer structure of Ti/$a$-C12A7/BaSnO$_3$/SrTiO$_3$ is clearly seen. The BaSnO$_3$ is heteroepitaxially grown on (001) SrTiO$_3$ substrate. Abrupt heterointerface between $a$-C12A7/BaSnO$_3$ is clearly seen.

**FIG. 3|** Typical transistor characteristics of the resultant BaSnO$_3$ TFT at room temperature [(a) transfer characteristics ($I_d$–$V_g$) at $V_d$=1 V, (b) $I_d^{0.5}$–$V_g$ plot, (c) output characteristics ($I_d$–$V_d$), and (d) field effect mobility ($\mu_{Hall}$)]. The on-to-off current ratio is ~10$^3$. The threshold voltage ($V_{th}$) is +5.5 V. The maximum field effect mobility ($\mu_{FE}$) is ~40 cm$^2$ V$^{-1}$ s$^{-1}$.

**FIG. 4|** Electric field modulated $S$ of the resultant BaSnO$_3$ TFT as a function of sheet carrier concentration at room temperature. An inflection point is seen at around (|$S$|, $n_s$)=(240 μV K$^{-1}$, 1.8×10$^{12}$ cm$^{-2}$). The slope of $S$–log $n_s$ plot is ~200 μV K$^{-1}$/decade (gray line) when $n_s$ exceeds 1.8×10$^{12}$ cm$^{-2}$. Judging from these data, there is a threshold of degenerate/non-degenerate semiconductor at around $n$=1.8×10$^{12}$ cm$^{-2}$.

Table I. Electrical properties of the La-doped BaSnO$_3$ films at room temperature.

| $x$ in Ba$_{1-x}$La$_x$SnO$_3$ | $\sigma$ (S cm$^{-1}$) | $n$ (cm$^{-3}$) | $\mu_{Hall}$ (cm$^2$ V$^{-1}$ s$^{-1}$) | $S$ (µV K$^{-1}$) |
|---|---|---|---|---|
| 0.001 | 0.00481 | 2.10×10$^{18}$ | 0.0143 | −253 |
| 0.004 | 1.32 | 8.80×10$^{18}$ | 0.935 | −243 |
| 0.0055 | 66.4 | 3.24×10$^{19}$ | 12.8 | −188 |
| 0.007 | 420 | 7.75×10$^{19}$ | 33.8 | −126 |
| 0.02 | 2620 | 2.66×10$^{20}$ | 61.6 | −57.2 |
| 0.05 | 6910 | 6.41×10$^{20}$ | 67.3 | −31.3 |
| 0.07 | 3350 | 4.76×10$^{20}$ | 43.9 | −38.4 |

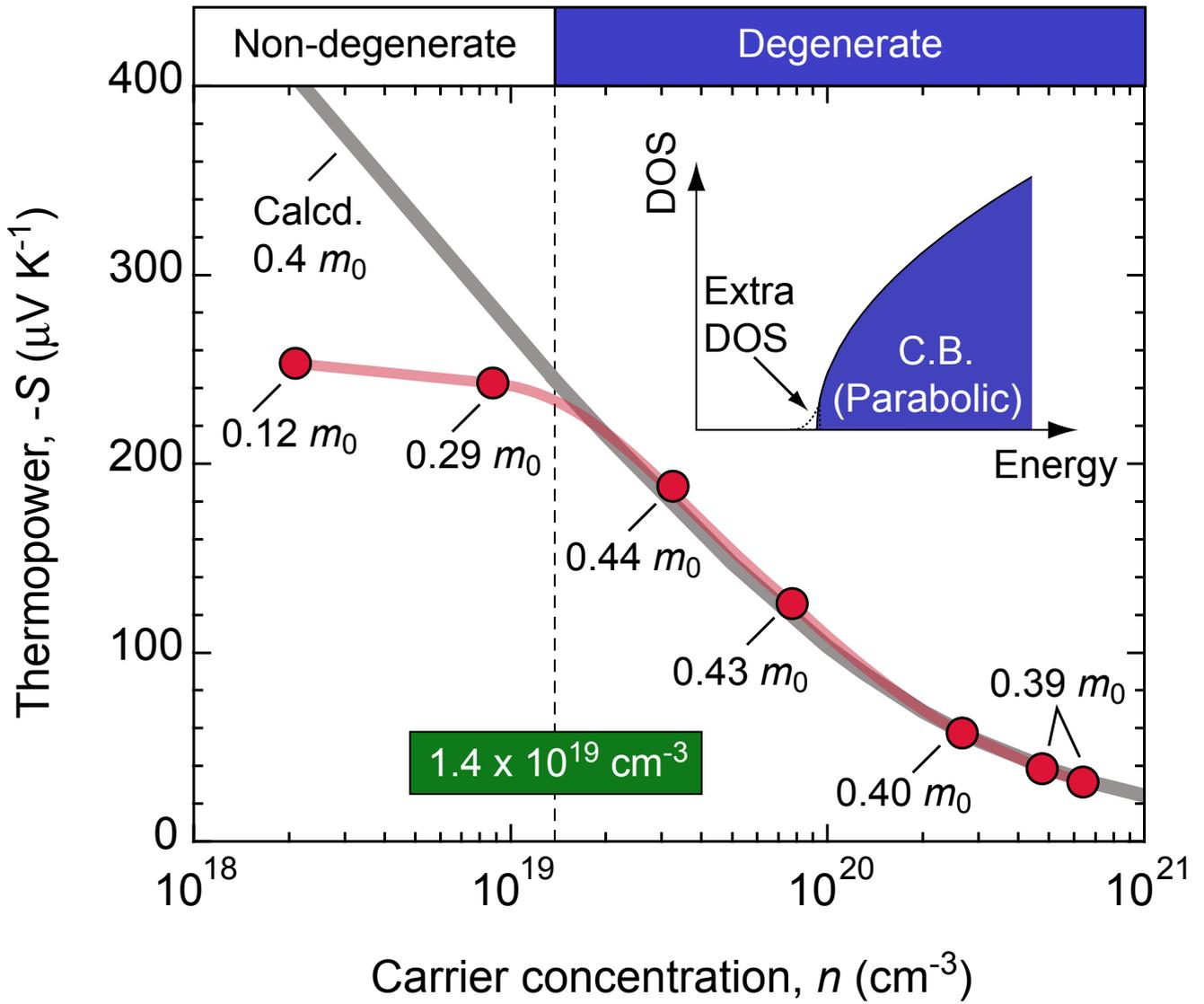

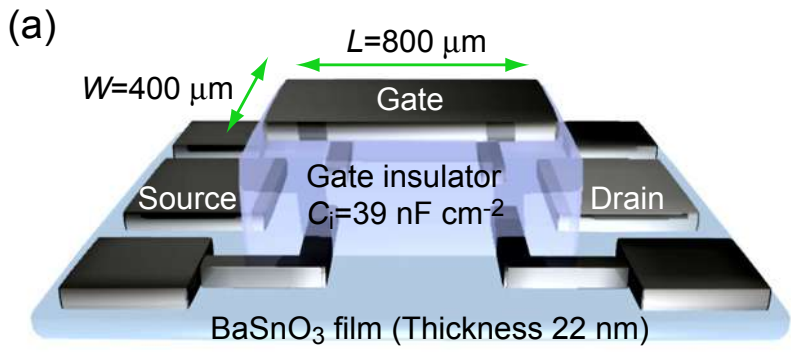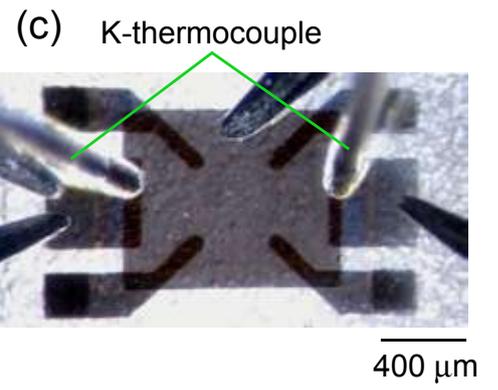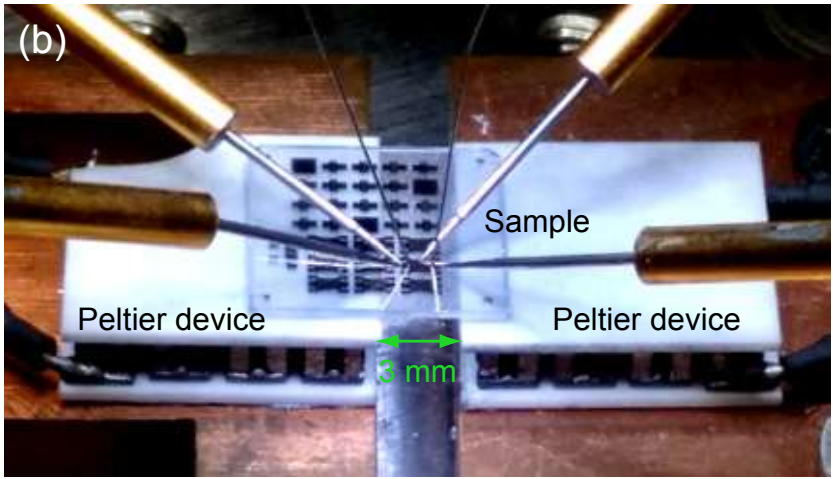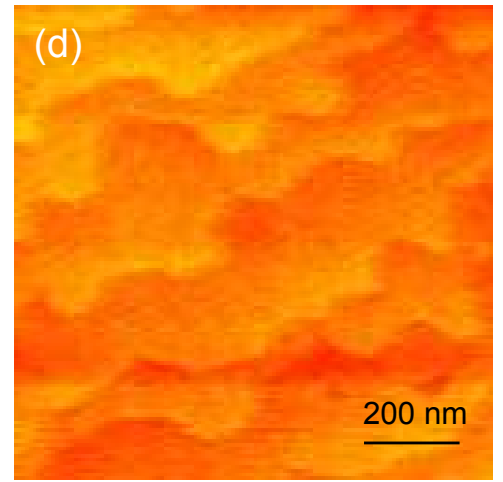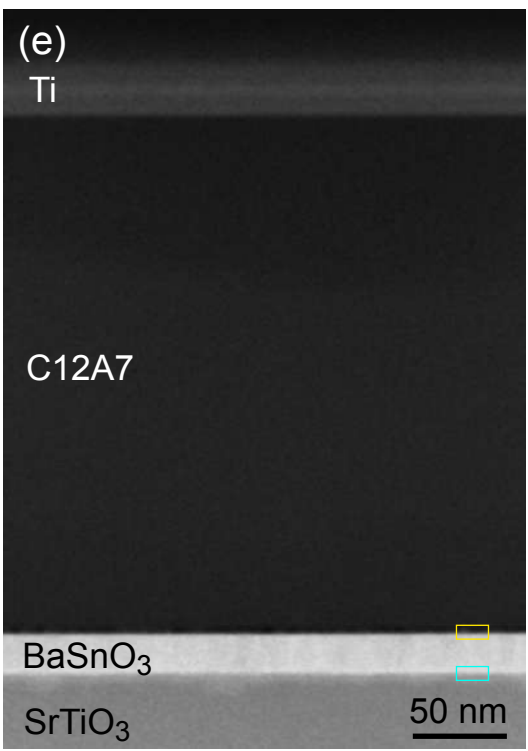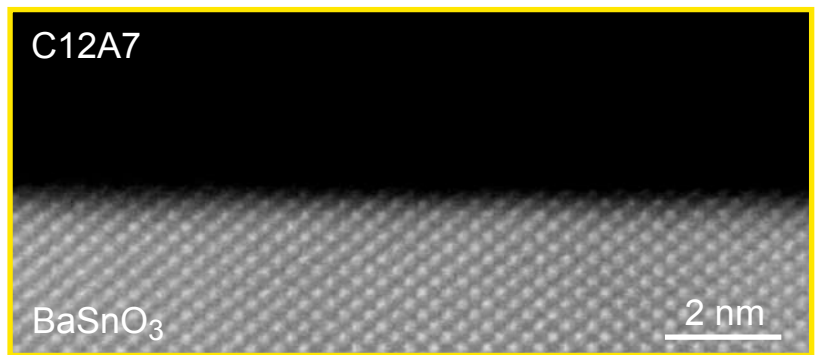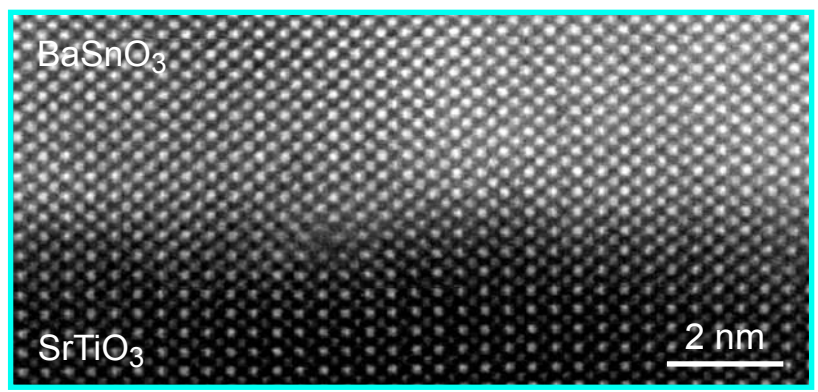

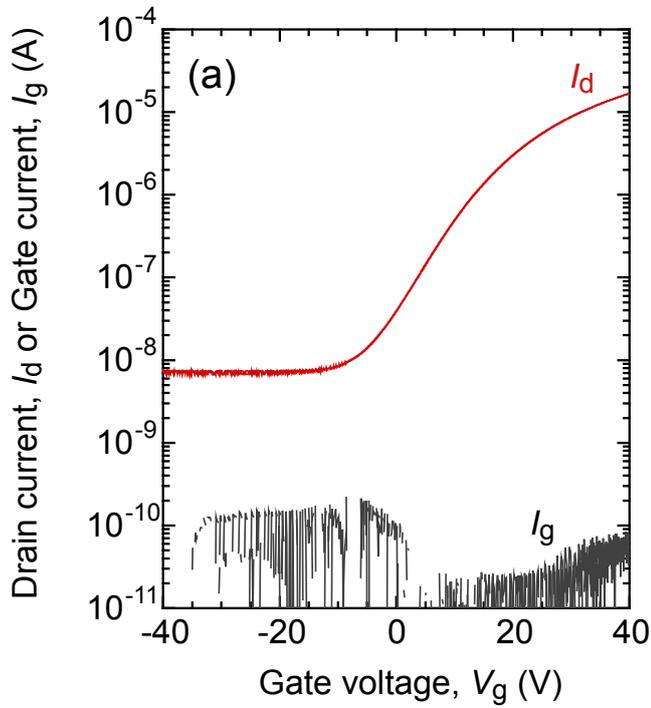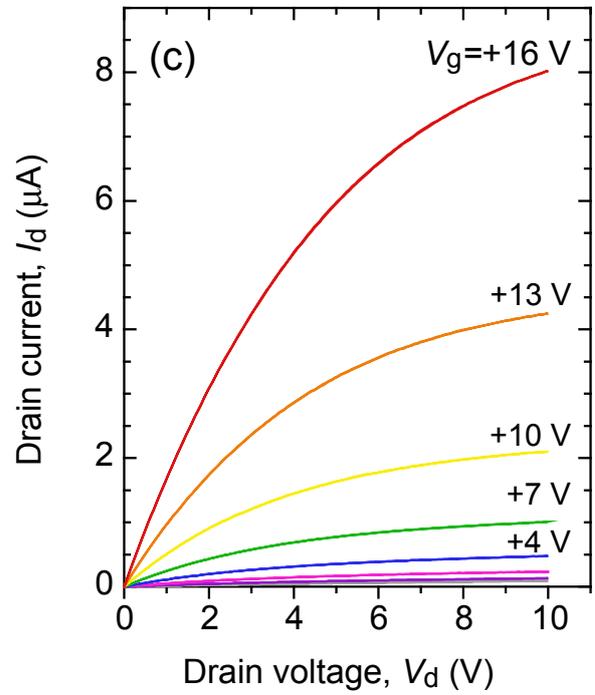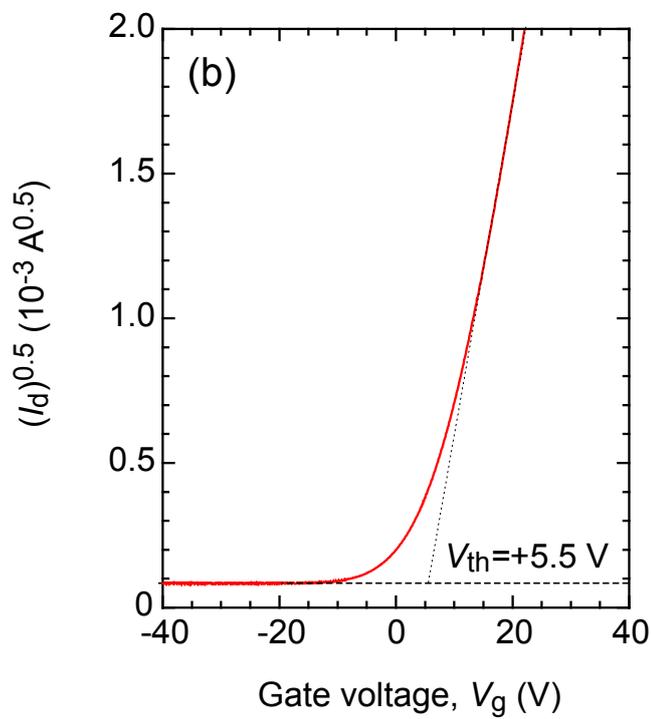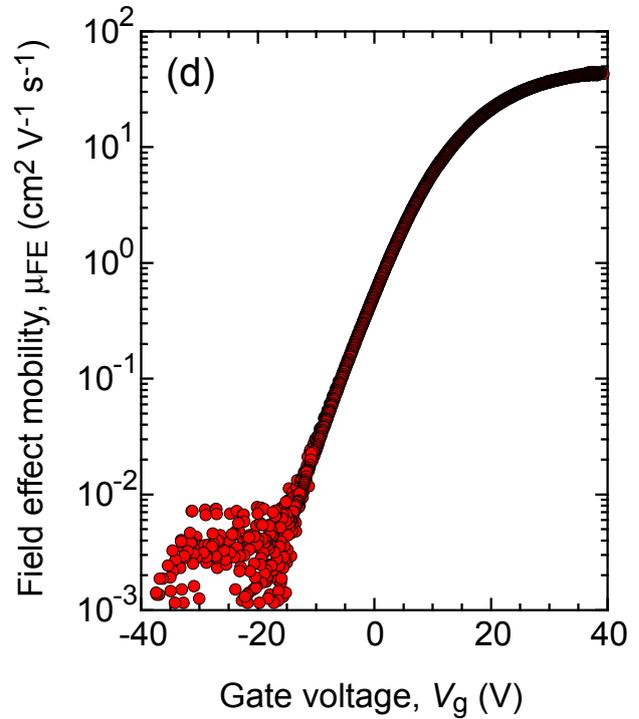

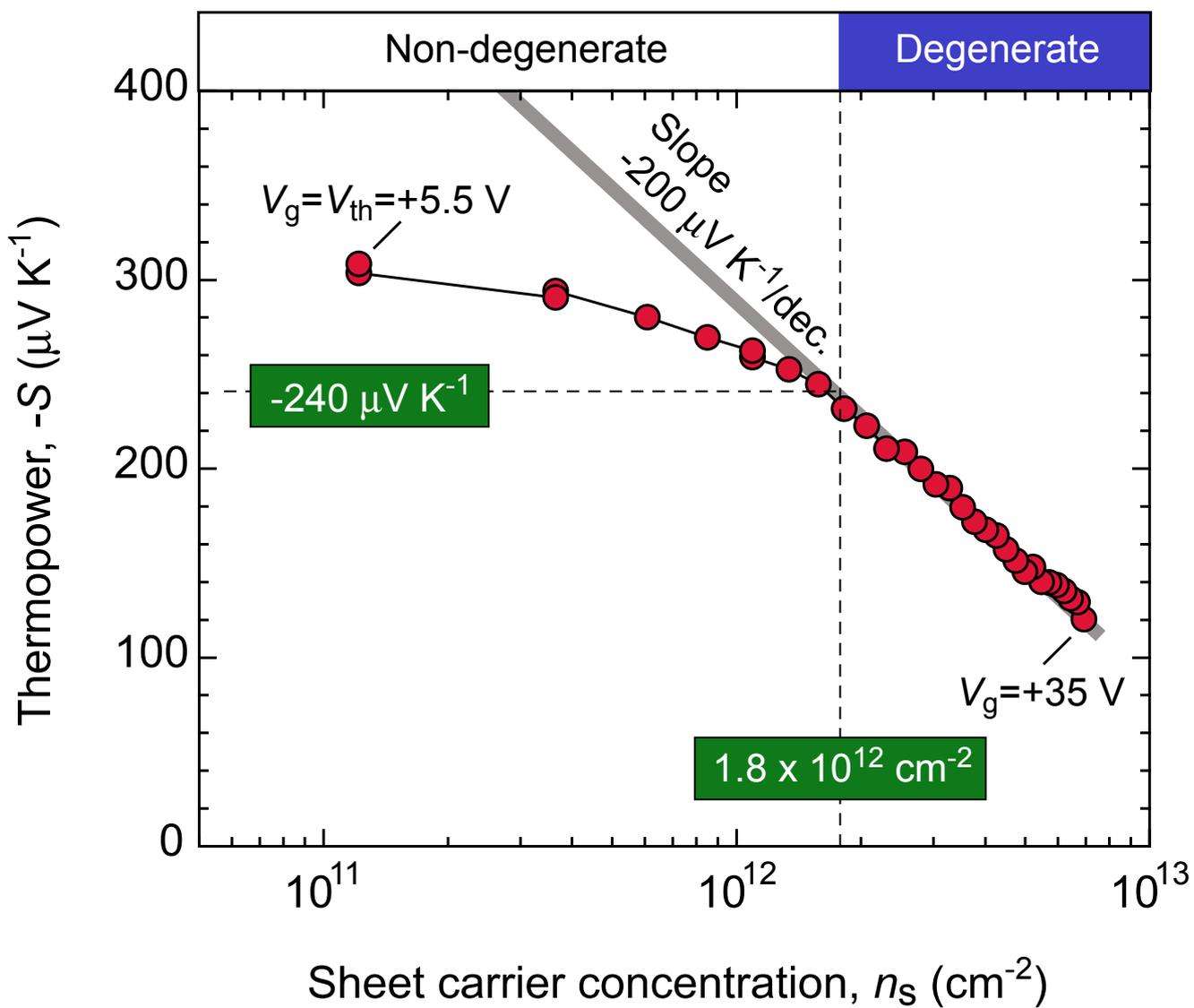